\documentclass[a4paper]{article}

\usepackage{INTERSPEECH2022}
\usepackage{multirow}
\usepackage{url}
\title{Uncertainty Calibration for Deep Audio Classifiers}
\name{Tong Ye$^{1,2\dagger}$, Shijing Si$^{1\dagger}$, Jianzong Wang$^{1\star}$\thanks{$^{\dagger}$ Equal contribution.}\thanks{$^\star$Corresponding author: Jianzong Wang, jzwang@188.com}, Ning Cheng$^1$, Jing Xiao$^1$}
\address{
  $^1$Ping An Technology (Shenzhen) Co., Ltd.\\
  $^2$University of Science and Technology of China}
\email{jzwang@188.com}

\begin{document}

\maketitle
\begin{abstract}
Although deep Neural Networks (DNNs) have achieved
tremendous success in audio classification tasks, their uncertainty calibration are still under-explored.
A well-calibrated
model should be accurate when it is certain about its prediction and indicate high
uncertainty when it is likely to be inaccurate. 
In this work, we investigate the uncertainty calibration for deep audio classifiers.
 In particular, we empirically study the performance of popular calibration methods: (i) Monte Carlo Dropout, (ii) ensemble, (iii) focal loss, and (iv) spectral-normalized Gaussian process (SNGP), on audio classification datasets. To this end, we evaluate (i–iv) for the tasks of environment sound and music genre classification. Results indicate that uncalibrated deep audio classifiers may be over-confident, and SNGP performs the best and is very efficient on the two datasets of this paper.
\end{abstract}

\section{Introduction}

Modern deep neural networks (DNNs) \cite{scarpiniti2021deep,si2021variational,bahmei2022cnn,si2022towards} have been widely utilized in many audio classification tasks such as multimedia search and retrieval, urban sound monitoring, bioacoustic monitoring, and audio captioning. For example, \cite{hershey2017cnn} has shown that fully connected multi-layered perceptron (MLP), AlexNet \cite{yu2016visualizing}, Inception \cite{szegedy2017inception}, and ResNet \cite{he2016deep} significantly outperforms raw features on the Audio Set \cite{gemmeke2017audio} for Acoustic Event Detection (AED) classification task.

Despite their extraordinary
performance, DNNs are often-criticized as being poorly calibrated and prone to be overconfident, thus leading to unsatisfied uncertainty estimation \cite{pleiss2017fairness,guo2017calibration,thulasidasan2019mixup}. The process of adapting deep learning's output to be consistent with the actual probability is
called uncertainty calibration, and has
drawn a growing attention in recent years \cite{minderer2021revisiting}.
In practical applications, miscalibrated probability estimates can be misleading in the sense
that the end user of these estimates has an incentive to mistrust (and therefore potentially misuse)
them \cite{fernando2021dynamically}. 


Many research have been devoted to calibrating deep models in machine learning, computer vision (CV) and natural language processing (NLP). 
\cite{guo2017calibration} explored with several classical calibration fixes and found that simple post-hoc methods like Temperature Scaling \cite{platt1999probabilistic} and Histogram Binning \cite{zadrozny2001obtaining} are significantly effective for DNNs. \cite{kull2019beyond,rahimi2020intra}
proposed to learn linear and non-linear
transformation functions to rescale the original output logits respectively. \cite{patel2021multi}
proposed a mutual information maximization-based binning
strategy to solve the severe sample-inefficiency issue in Histogram Binning.
\cite{muller2019does} showed
that training models using the standard CE loss with label smoothing, instead of one-hot
labels, has a very favourable effect on model calibration. \cite{mukhoti2020calibrating} proposed to improve uncertainty calibration by replacing the conventionally used CE loss with the focal loss proposed in \cite{lin2017focal} when training DNNs.

\begin{figure}[htp]
    \centering
    \includegraphics[width=\linewidth]{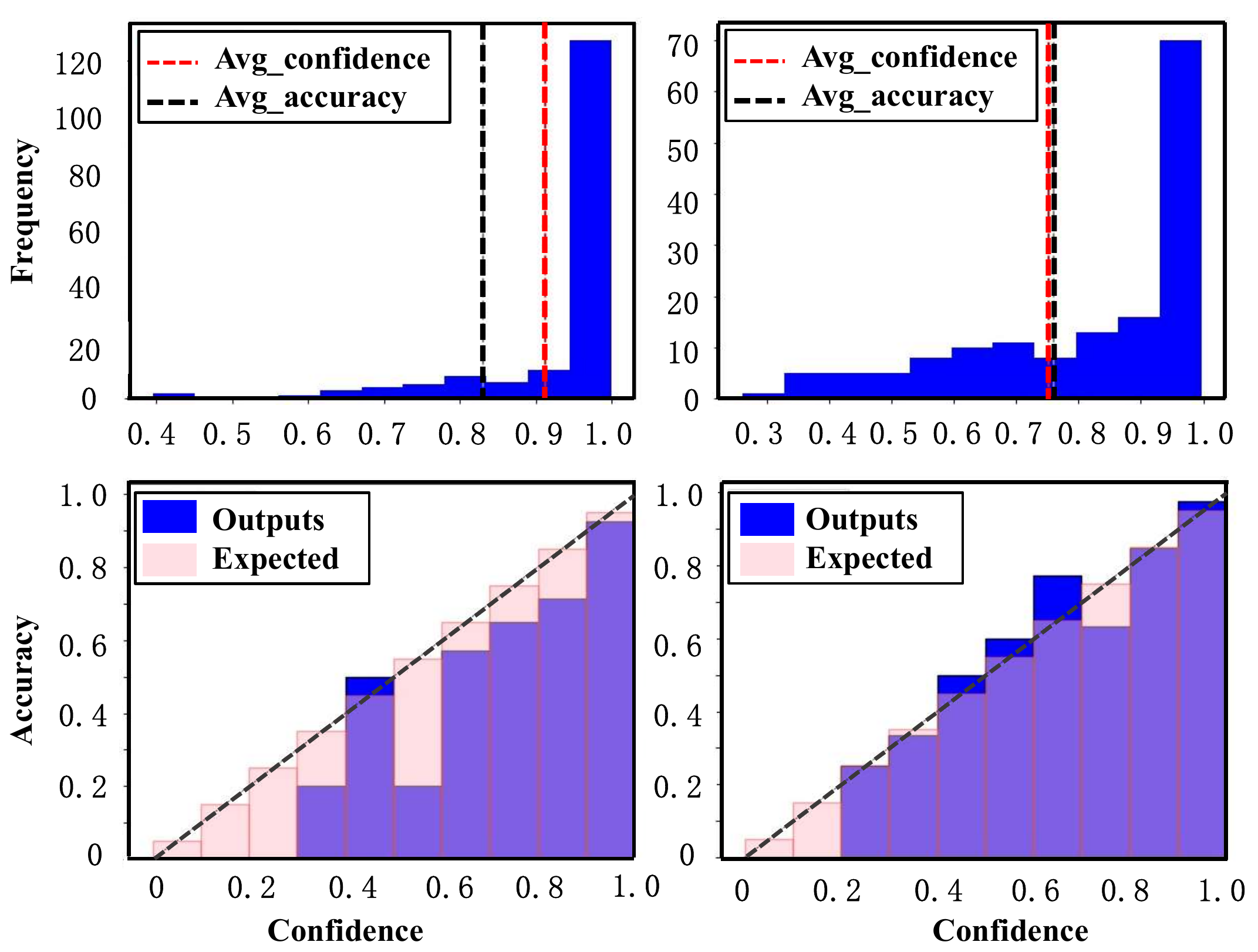}
    \caption{Confidence histograms (top) and reliability diagrams (bottom) for a base ResNet-50 audio classifier (left) and a calibrated method (right) on the ESC-50 dataset.}
    \label{fig:calib}
\end{figure}

However, model calibration for audio classifiers is still under-explored.
Our goal is not only to understand whether deep audio classifiers are miscalibrated, but also to study what methods can alleviate this problem. As shown in Fig. \ref{fig:calib}, through an empirical study we find that audio classifiers using ResNet-50 is over-confident. In the topleft plot, the average confidence of all samples is about 0.91, but the accuracy is only 0.84. The topright shows the performance of a classifier calibrated by focal loss. Though the accuracy decreases to around 0.75, the model's confidence is consistent to its accuracy. 
We argue that it is not trivial to transfer expertise in CV and NLP areas to audio classification, due to the difference between various modalities.
we compare various calibration methods on three popular network architectures, Inception, ResNet and DenseNet, and examine their performances on two audio classification datasets.

Our contributions can be summarized as follows:
\begin{itemize}
    \item We verify the existence of miscalibration of deep classifiers for audio datasets, which can raise the community's awareness of this uncertainty calibration problem.
    \item We empirically examine the performance of various calibration methods for audio classifiers, with SNGP performs the best and is efficient. 
\end{itemize}

\section{Background}

\subsection{Definition}

Here we present some basic concepts for model calibration. In this paper, we consider the multi-classification problem for audio data, where we observe an audio (or its features) $X$ and predict a categorical variable $Y \in \{1, 2, \ldots, K\}$. A predictor $f$ as a function that maps every input instance $X$ to a categorical distribution over $K$ labels, represented using a
vector $f(X)$ belonging to the $(K-1)$-dimensional simplex
$\Delta = \{p \in [0, 1]^{K}|\sum_{y=1}^{K}p_{y} = 1\}$. 

Intuitively, a model $f$ is well-calibrated if its output truthfully quantifies the predictive uncertainty. 
For example, if we take all data points $x$ for which the model predicts $[f(x)]_{y} = 0.4$, we expect 40\% of them to indeed have the label $y$. Formally, the model $f$ is said to be calibrated if \cite{brocker2009reliability}
\begin{equation}\label{eq:defi}
    \forall p\in\Delta: P(Y = y | f(X) = p) = p_{y}.
\end{equation}


The most common measure of the degree of miscalibration is the Expected Calibration Error (ECE),
which computes the expected disagreement between confidence and accuracy. Typically we first bucket the predictions into $m$ (usually $m = 10$) bins $B_1, \ldots, B_{m}$ based on their top predicted
probability, and then takes the expectation over these buckets. Namely, if we are given a set of $n$
i.i.d. samples $(x_{1}, y_{1}), \ldots, (x_{n}, y_{n})$,
then we assign each $j \in \{1, \ldots, n\}$
to a bucket $B_i$ based on $\max f(x_j)$.
Consequently, we compute in each bucket $B_i$ the 
\begin{equation}\label{eq:conf}
  \text{confidence}(B_{i}) = \frac{1}{|B_{i}|}\sum_{j\in~B_{i}}\max f(x_j),\vspace{-0.5cm}
\end{equation}
\begin{equation}\label{eq:acc}
  \text{accuracy}(B_{i}) = \frac{1}{|B_{i}|}\sum_{j\in~B_{i}}1[y_{j} \in \arg\max f(x_{j})],
\end{equation}
where the $1[y_{j} \in \arg\max f(x_{j})]$ is an indicator function, taking $1$ if $y_{j} \in \arg\max f(x_{j})$ otherwise $0$. 
Finally, ECE is evaluated by taking the expectation over the bins
\begin{equation}\label{eq:ece}
  \hat{\text{ECE}} = \sum_{i=1}^{m}\frac{|B_{i}|}{n}|\text{accuracy}(B_{i}) - \text{confidence}(B_{i})|.
\end{equation}

\subsection{Popular Calibration Methods}
Here we summarize popular uncertainty calibration methods widely used in the literature.

\noindent\textbf{Monte Carlo Dropout}
\cite{gal2016dropout} proposes a way to approximate Bayesian inference by employing dropout and generates a predictive distribution after a number of forward passes. Monte Carlo (MC) Dropout \cite{penha2021calibration,cohen2021not} is easy to use, and has zero memory overhead compared to a single model.
Unfortunately, it requires multiple forward passes which also result in a substantial obstacle given the prevalence of BERT and other large transformer architectures \cite{durasov2021masksembles}.

\noindent\textbf{Ensemble}
 method casts a set of models under the same architecture with different parameter initialization or other perturbations, encouraging independent member predictions. At test time, the ensemble prediction is the average
of soft-max outputs of multiple individually trained models
to evaluate the final accuracy. Independent trained identical
models create diversity in ensembles due to differences in
model initialization and mini-batch orderings \cite{lakshminarayanan2017simple},
which results in different local optimal solutions.
\cite{zhang20mix} proposed the Mix-n-Match calibration strategies which achieves remarkably better data-efficiency and expressive power while provably maintaining the classification accuracy of the original classifier. However, ensemble methods are
parameter-efficient but still require multiple forward passes from the model, which consumes larger computing resources than other methods.

\noindent\textbf{Focal loss}
 is originally proposed to address the class imbalance problem in object detection \cite{lin2017focal}. It
reshapes the standard CE loss through weighting loss components of all samples according to how
well the model fits them. Therefore it focuses on fitting hard samples and prevents the easy samples from
overwhelming the training procedure.
\cite{mukhoti2020calibrating} verified the effectiveness of focal loss for uncertainty calibration.
\cite{charoenphakdee2021focal} studied how to recover the true class-posterior probability from the outputs of the focal risk minimizer.

\noindent\textbf{Spectral-normalized Neural Gaussian Process (SNGP)}
This method employs a Gaussian process,
boosting the model’s ability to properly quantify the distance of a testing example from the training data manifold and enable a DNN to achieve high-quality uncertainty estimation \cite{liu2020simple}.
Specifically, on top of modern DNNs, it adds
a weight normalization step during training and replacing the output layer with a
Gaussian Process.

\section{Experiments}
We implement various calibration methods in Python. Our code is publicly available at a github repository\footnote{\url{https://github.com/shijing001/Unicertainty_calibration_audio_classifiers}}.
\subsection{Datasets}
We conduct extensive experiments on two commonly used datasets: ESC-50 and GTZAN. Details on these two datasets are presented as follows.

\noindent\textbf{ESC-50} is a collection of short environmental recordings available in a unified format (5-second-long clips, 44.1 kHz, single channel, Ogg Vorbis compressed @ 192 kbit/s). It consists of a labeled set of 2000 environmental recordings (50 classes, 40 clips per class). We split the whole dataset into training, validation and testing sets in the ratio of 8:1:1, while keeeping the labels balanced.

\noindent\textbf{GTZAN}\cite{tzanetakis2002musical} The GTZAN dataset consists of 1000
music clips each of length 30s. There are 10 distinct genre classes. The music clips are sampled at a rate of 22.5 kHz. There is no official training and validation split of the dataset. Therefore we split the whole dataset into training, validation and testing sets in the ratio of 6:2:2, while keeeping the labels balanced.

\subsection{Preprocessing}

The input audio signal is re-sampled to 22.5 kHz at the pre-processing step. Re-sampling is applied to reduce dimensionality of the input
signal. In addition, every sample is padded with zeros to guarantee uniformity in input data. Each audio is transformed as a 2-dimensional feature map representing frequencies
with respect to time \cite{ajmera2011text}.
Since mel-spectrograms with different window sizes 
and hop lengths in each channel yield varied classification performance. 
The mel-spectrograms were obtained using 128 mel bins 
and then log scaled. For ESC-50, we use the input 
of size (128, 250), whereas, for GTZAN, we use the input of size (128, 1500).

\subsection{Experimental configuration}
A well-calibrated deep learning model should: 1.)
produce confidence scores close to its accuracy; and
2.) exhibit higher uncertainty on inputs far
away from training data. To empirically evaluate the performance of calibration methods,
our experiments are divided into two parts: in-distribution calibration and out-of-distribution detection. In-distribution calibration
measures how well a model’s predicted confidence aligns
with observed accuracy.
Out-of-distribution detection measures the ability of a model to reject OOD inputs.

For the in-distribution calibration, we train classifiers with various methods, and take the output of softmax as predicted probabilities, and then evaluate the ECE scores. For evaluating out-of-distribution detection,
we conduct our experiments in the similar approach as introduced by \cite{liang2018enhancing}. In these experiments, a neural network is first trained on some ESC-50 data, which represents
the in-distribution examples. Out-of-distribution examples
are represented by music audio examples from GTZAN that contain classes different from those found in the in-distribution dataset. For each sample in the in-distribution test set, and each out-of-distribution example, a confidence score is produced, which will be used to predict which distribution the
samples come from. Finally, several different evaluation
metrics are used to measure and
compare how well different confidence estimation methods
can separate the two distributions.



\subsection{Deep Learning models}
We employ three popular CNN architectures as the backbone in our experiments: Inception, ResNet and DenseNet.

\noindent\textbf{Inception} An Inception Layer \cite{szegedy2016rethinking} is a combination of
all the layers namely, $1 \times 1$ Convolutional layer, $3\times 3$ Convolutional layer, $5\times 5$ Convolutional layers with their output filter banks concatenated into a single output vector. 
Here we used Inception-V3 backbone.

\noindent\textbf{ResNet}
\cite{he2016deep} The residual block has two $3 \times 3$ convolutional layers with the same number of output channels. Each convolutional layer is followed by a batch normalization layer and a ReLU activation. A skip connection is added which skips these two convolution operations and adds the input directly
before the final ReLU activation. 
Here we used ResNet-50 backbone.

\noindent\textbf{DenseNet}\cite{huang2017densely}
Dense Convolutional Network
(DenseNet), connects each layer to every other
layer in a feed-forward fashion. For each layer, the
feature-maps of all preceding layers are used as inputs,
and its own feature-maps are used as inputs into all
subsequent layers. Traditional convolutional networks
with $L$ layers have $L$ connections one between each layer and its subsequent layer a dense network has
$L(L+1)/2$ direct connections.
We used DenseNet-201 backbone for the experiments.

\subsection{Calibration Methods}
Here are some details on how we implement popular calibration methods.

\noindent\textbf{Focal loss} makes the model focus on hard training examples, paying less attention to easy examples. In this experiment, we set the tuning parameters $\alpha = 0.25$ and $\gamma = 2$ for focal loss.

\noindent\textbf{MC Dropout}: We implement the dropout with 10 dropout samples for all CNN layers with probability $0.1$.

\noindent\textbf{Ensemble}
We trained $M = 5$ independent models to predict
audio classification scores, using the same architecture, with different initialization values. At test time, the ensemble prediction is the average of soft-max outputs of these $M$ individually trained models to evaluate the final accuracy. 

\noindent\textbf{SNGP}
Following \cite{liu2020simple}, we implement SNGP methods for three network architectures, and employ Laplace approximation for inference.

\begin{table}[]
\caption{Accuracy and ECE for In-distribution calibration of the base architecture (no calibration) and four calibration methods (focal loss, MC dropout, Ensemble and SNGP) on two datasets\label{tab:in.dist}}
\begin{tabular}{llllll}
\hline
                                        Archit.      &    Method    & \multicolumn{2}{l}{ESC-50} & \multicolumn{2}{l}{GTZAN} \\
                                            &           & Acc$\uparrow$          & ECE$\downarrow$         & Acc$\uparrow$         & ECE$\downarrow$          \\ \hline
\multicolumn{1}{c}{\multirow{5}{*}{ResNet}} & +base     & 0.835        & 0.106       & 0.734       & 0.195       \\
\multicolumn{1}{c}{}                        & +focal    & 0.765        & 0.049       & 0.643       & 0.127       \\
\multicolumn{1}{c}{}                        & +Dropout  & 0.830        & 0.093       & 0.764       & 0.121       \\
\multicolumn{1}{c}{}                        & +Ensemble & 0.831        & 0.091       & 0.738       & 0.184       \\
\multicolumn{1}{c}{}                        & +SNGP     & \textbf{0.845}        & \textbf{0.048}       & \textbf{0.784}       & \textbf{0.069}       \\ \hline
\multirow{5}{*}{DenseNet}                   & +base      & 0.905        & 0.059       & 0.829       & 0.077       \\
                                            & +focal    & 0.886        & 0.055       & 0.822       & 0.057       \\
                                            & +Dropout  & 0.915        & 0.053       & \textbf{0.849}      & \textbf{0.054}       \\
                                            & +Ensemble & 0.895        & 0.051       & 0.844       & 0.071       \\
                                            & +SNGP     & \textbf{0.930}        & \textbf{0.034}       & 0.839      & 0.075       \\ \hline
\multirow{5}{*}{Inception}                  & +base      & 0.715        & 0.138       & 0.754       & 0.158       \\
                                            & +focal    & 0.644        & 0.106       & 0.758       & \textbf{0.054}       \\
                                            & +Dropout  & 0.720        & 0.073       & 0.748       & 0.121       \\
                                            & +Ensemble & 0.728        & 0.122       & 0.750       & 0.149       \\
                                            & +SNGP     & \textbf{0.785}        & \textbf{0.054}       & \textbf{0.779}       & 0.086      \\ \hline
\end{tabular}
\vspace{-0.5cm}
\end{table}

\begin{figure*}[ht!]
    \centering
    \includegraphics[width=.95\linewidth]{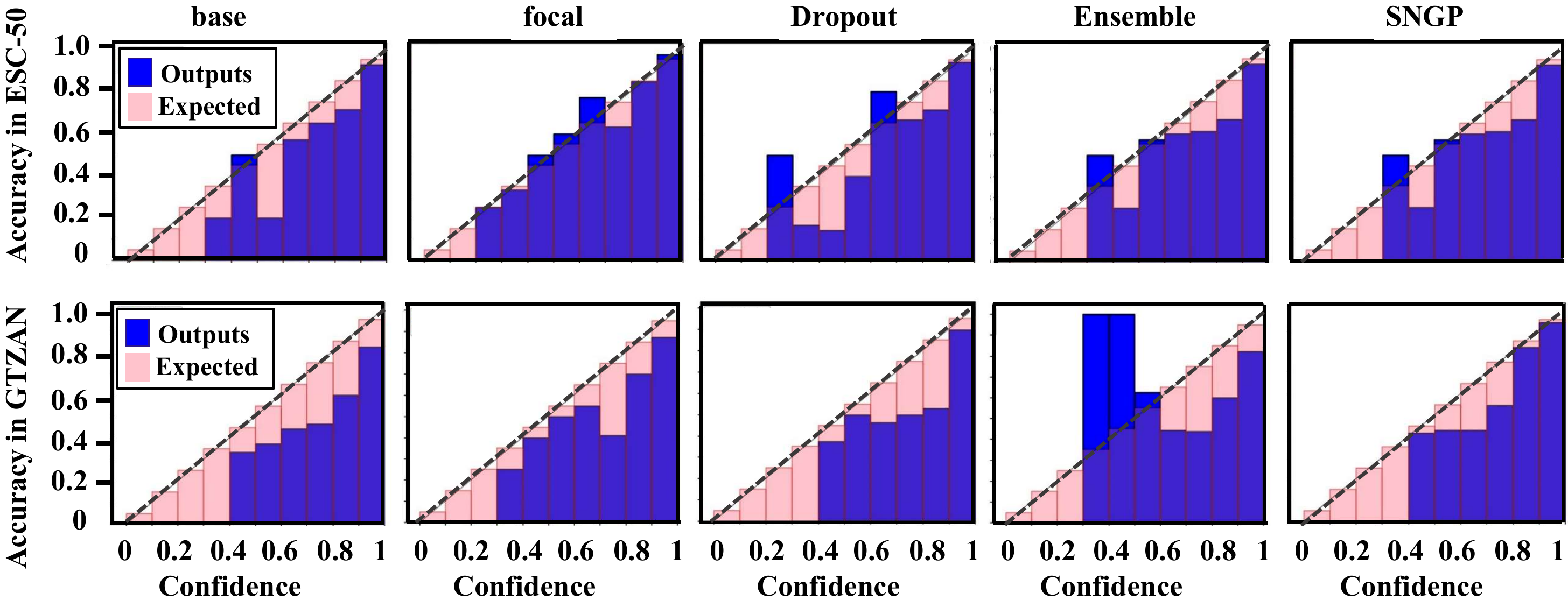}
    \caption{Reliability diagrams of the ResNet-50 architecture and four calibration methods (focal loss, MC dropout, Ensemble and SNGP) on two datasets: ESC-50 (top row) and GTZAN (bottom row). Less gap between the expected (pink bars) and the output (blue bars) means better performance.}
    \label{fig:in.dist}
    \vspace{-.5cm}
\end{figure*}

\subsection{Evaluation Metrics}

\noindent\textbf{ECE:}\cite{naeini2015obtaining} is used to evaluate calibration metric from in-distribution classification. We group all samples into $m = 10$ equally interval bins with respect to their confidence scores, then calculating the expected difference
between the accuracy and average confidence, shown in Eq. \eqref{eq:ece}. Smaller ECE scores means better performance.

\noindent\textbf{AUROC:} measures the Area Under the Receiver Operating
Characteristic curve. It can be interpreted as
the probability that a positive example (in-distribution) will have a higher detection score than a negative example (out-of-distribution).

\noindent\textbf{AUPR:} measures the Area Under the Precision-Recall (PR)
curve. The PR curve is made by plotting $precision =
TP/(TP + FP)$ versus $recall = TP/(TP + FN)$. In our
tests, AUPR indicates that
out-of-distribution examples are used as the positive class. Both AUROC and AUPR are used to evaluate the performance of out-of-distribution detection, and larger values meaning better performance.

\subsection{Results and Analysis}
Here we present the results and analysis of our experiments.
 
\begin{table}[]
\caption{Performance of the base architecture and four calibration methods (focal loss, MC dropout, Ensemble and SNGP) for out-of-distribution (OOD) detection\label{tab:ood}}
\begin{tabular}{llll}
\hline
                                      Archit.      &    Method       & AUROC$\uparrow$ & AUPR$\uparrow$  \\ \hline
\multicolumn{1}{c}{\multirow{5}{*}{ResNet}} & +base     & 0.828 & 0.848 \\
\multicolumn{1}{c}{}                        & +focal    & 0.756 & 0.788 \\
\multicolumn{1}{c}{}                        & +Dropout  & 0.834 & 0.858 \\
\multicolumn{1}{c}{}                        & +Ensemble & 0.835 & 0.853 \\
\multicolumn{1}{c}{}                        & +SNGP     & \textbf{0.849} & \textbf{0.881} \\ \hline
\multirow{5}{*}{DenseNet}                   & +base      & 0.879 & 0.894 \\
                                            & +focal    & 0.885 & 0.906 \\
                                            & +Dropout  & 0.878 & 0.893 \\
                                            & +Ensemble & 0.885 & 0.900 \\
                                            & +SNGP     & \textbf{0.928} & \textbf{0.944} \\ \hline
\multirow{5}{*}{Inception}                  & +base      & 0.713 & 0.763 \\
                                            & +focal    & 0.643 & 0.661 \\
                                            & +Dropout  & 0.724 & 0.760 \\
                                            & +Ensemble & 0.733 & 0.778 \\
                                            & +SNGP     & \textbf{0.788} & \textbf{0.811} \\ \hline
\end{tabular}
\vspace{-.5cm}
\end{table}

\noindent\textbf{In-distribution calibration}
We begin by considering ECE on two datasets: ESC-50 and GTZAN. Table \ref{tab:in.dist}
shows in-distribution ECE and accuracy of the three base architectures (Inception-V3, ResNet-50, and DenseNet-201) and four calibration methods.
As shown in Table \ref{tab:in.dist}, for predictive accuracy, SNGP consistently performs the best for both datasets across three network architectures. For calibration error (ECE), SNGP clearly outperforms the
other approaches on ESC-50 dataset and is also competitive on GTZAN dataset. The performance of other methods vary significantly across different architectures and datasets, but are significantly better than the uncalibrated base model in terms of ECE. Among three architectures, DenseNet-201 achieves better accuracy and ECE than ResNet-50 and Inception-V3. This is mainly because it has much more layers (201) than others.
Therefore, network architecture also affects the performance of calibration methods.

Figure \ref{fig:in.dist} displays the reliability diagrams of the base ResNet-50 (no calibration) and four calibration methods on the two datasets: ESC-50 (top row) and GTZAN (bottom row). On each plot, less gap between the output bars (blue) and the expected bars (pink) means better performance. From this figure, both focal loss and SNGP yields less gap than other methods.

\begin{figure}[htp]
    \centering
    \includegraphics[width=\linewidth]{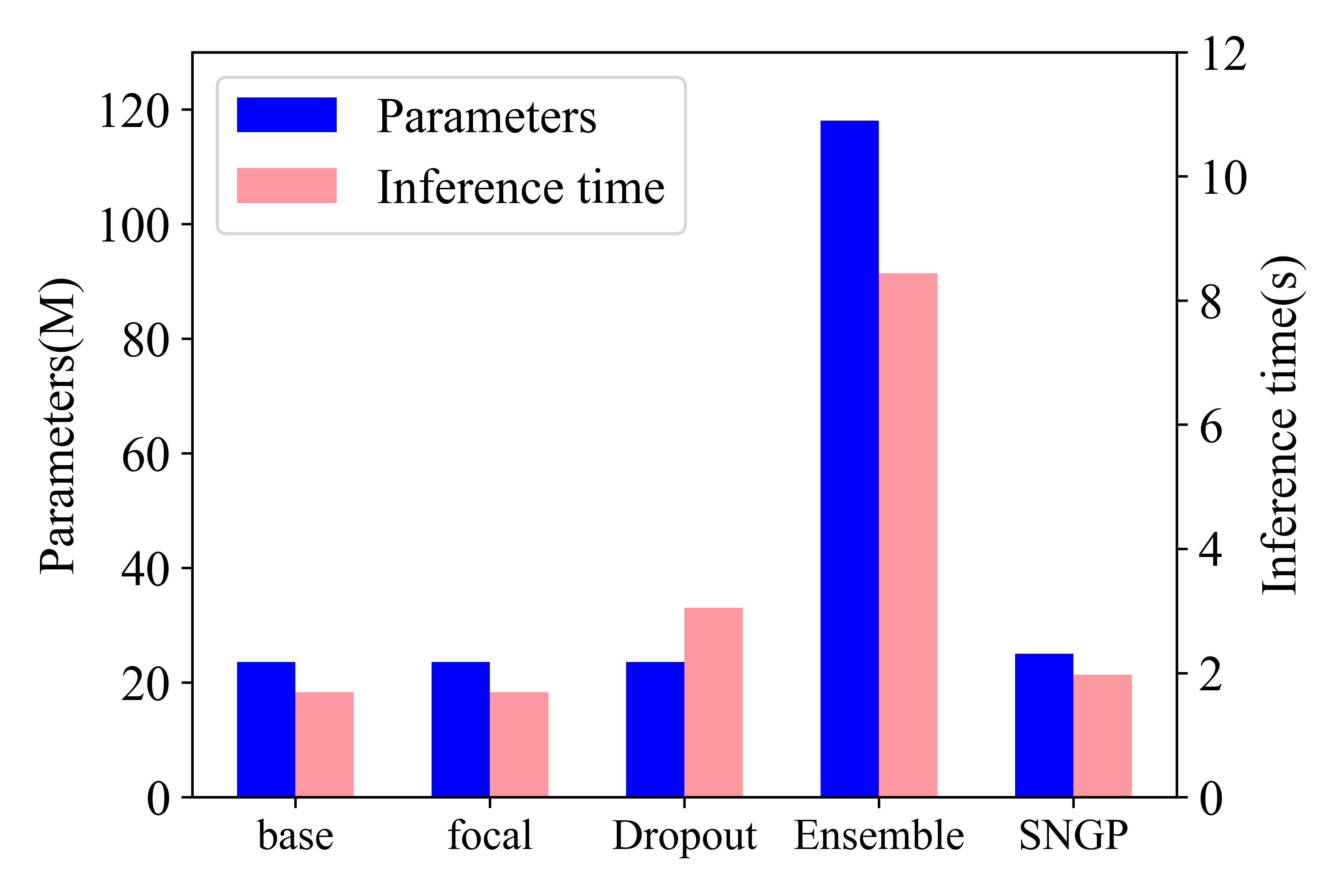}
    \caption{The number of parameters (by Million) and inference time (by milli-seconds) in five methods.}
    \label{fig:params}
    \vspace{-.5cm}
\end{figure}

\noindent\textbf{Out-of-distribution detection}
To evaluate how suitable the learned confidence estimates
are for separating in- and out-of-distribution examples, we conduct out-of-distribution detection and compare the performance of various calibration methods.
Table \ref{tab:ood} exhibits the performance of three base architectures and four methods on out-of-distribution detection. For this task,
SNGP performs the best across all datasets and architectures, followed by the ensemble method.
In general, calibration methods perform no worse than the uncalibrated base method, except the focal loss, which performs the worst on ResNet-50 and Inception. This means that focal loss could lead to a classifier bad at discriminating in- and out-of-distribution samples.

\noindent\textbf{Computing Efficiency} To compare the efficiency of calibration methods,
Figure \ref{fig:params} shows the the number of parameters (by millions) and inference time for one single sample (by milli-seconds). This figure takes ResNet-50 as the base model. The ensemble method has largest number of parameters, almost 5 times more than the others. In terms of inference time, ensemble consumes the most, followed by MC dropout. SNGP and focal loss are very efficient, close to the uncalibrated baseline.

In summary, SNGP method performs the best on uncertainty calibration and also very efficient to implement. Through focal loss can produce good calibration, it performs bad at out-of-distribution detection. Ensemble performs well at out-of-distribution detection, but it is not efficient.

\section{Conclusion}

Audio classification has witnessed rapid improvement as an increasing number of deep learning models are deployed.
However, calibration for audio classifiers is still under-explored. In this work, we investigate the performance of calibration methods for deep audio classifiers, verifying the effectiveness of SNGP and ensemble to audio classifiers. This will raise this community's awareness to the uncertainty calibration issue.

\section{Acknowledgements}

This paper is supported by the Key Research and Development Program of Guangdong Province under grant No.2021B0101400003. Corresponding author is Jianzong
Wang from Ping An Technology (Shenzhen) Co., Ltd (jzwang@188.com).

\newpage

\bibliographystyle{IEEEtran}

\bibliography{mybib}

\begin{thebibliography}{10}
\providecommand{\url}[1]{#1}
\csname url@samestyle\endcsname
\providecommand{\newblock}{\relax}
\providecommand{\bibinfo}[2]{#2}
\providecommand{\BIBentrySTDinterwordspacing}{\spaceskip=0pt\relax}
\providecommand{\BIBentryALTinterwordstretchfactor}{4}
\providecommand{\BIBentryALTinterwordspacing}{\spaceskip=\fontdimen2\font plus
\BIBentryALTinterwordstretchfactor\fontdimen3\font minus
  \fontdimen4\font\relax}
\providecommand{\BIBforeignlanguage}[2]{{%
\expandafter\ifx\csname l@#1\endcsname\relax
\typeout{** WARNING: IEEEtran.bst: No hyphenation pattern has been}%
\typeout{** loaded for the language `#1'. Using the pattern for}%
\typeout{** the default language instead.}%
\else
\language=\csname l@#1\endcsname
\fi
#2}}
\providecommand{\BIBdecl}{\relax}
\BIBdecl

\bibitem{scarpiniti2021deep}
M.~Scarpiniti, D.~Comminiello, A.~Uncini, and Y.-C. Lee, ``Deep recurrent
  neural networks for audio classification in construction sites,'' in
  \emph{2020 28th European Signal Processing Conference (EUSIPCO)}.\hskip 1em
  plus 0.5em minus 0.4em\relax IEEE, 2021, pp. 810--814.

\bibitem{si2021variational}
S.~Si, J.~Wang, H.~Sun, J.~Wu, C.~Zhang, X.~Qu, N.~Cheng, L.~Chen, and J.~Xiao,
  ``Variational information bottleneck for effective low-resource audio
  classification,'' in \emph{Proceedings of the Annual Conference of the
  International Speech Communication Association, INTERSPEECH}, 2021, p.~31.

\bibitem{bahmei2022cnn}
B.~Bahmei, E.~Birmingham, and S.~Arzanpour, ``Cnn-rnn and data augmentation
  using deep convolutional generative adversarial network for environmental
  sound classification,'' \emph{IEEE Signal Processing Letters}, 2022.

\bibitem{si2022towards}
S.~Si, J.~Wang, J.~Peng, and J.~Xiao, ``Towards speaker age estimation with
  label distribution learning,'' in \emph{ICASSP 2022-2022 IEEE International
  Conference on Acoustics, Speech and Signal Processing (ICASSP)}.\hskip 1em
  plus 0.5em minus 0.4em\relax IEEE, 2022, pp. 4618--4622.

\bibitem{hershey2017cnn}
S.~Hershey, S.~Chaudhuri, D.~P. Ellis, J.~F. Gemmeke, A.~Jansen, R.~C. Moore,
  M.~Plakal, D.~Platt, R.~A. Saurous, B.~Seybold \emph{et~al.}, ``Cnn
  architectures for large-scale audio classification,'' in \emph{ICASSP}.\hskip
  1em plus 0.5em minus 0.4em\relax IEEE, 2017, pp. 131--135.

\bibitem{yu2016visualizing}
W.~Yu, K.~Yang, Y.~Bai, T.~Xiao, H.~Yao, and Y.~Rui, ``Visualizing and
  comparing alexnet and vgg using deconvolutional layers,'' in \emph{ICML},
  2016.

\bibitem{szegedy2017inception}
C.~Szegedy, S.~Ioffe, V.~Vanhoucke, and A.~A. Alemi, ``Inception-v4,
  inception-resnet and the impact of residual connections on learning,'' in
  \emph{Thirty-first AAAI conference on artificial intelligence}, 2017.

\bibitem{he2016deep}
K.~He, X.~Zhang, S.~Ren, and J.~Sun, ``Deep residual learning for image
  recognition,'' in \emph{CVPR}, 2016, pp. 770--778.

\bibitem{gemmeke2017audio}
J.~F. Gemmeke, D.~P. Ellis, D.~Freedman, A.~Jansen, W.~Lawrence, R.~C. Moore,
  M.~Plakal, and M.~Ritter, ``Audio set: An ontology and human-labeled dataset
  for audio events,'' in \emph{ICASSP}.\hskip 1em plus 0.5em minus 0.4em\relax
  IEEE, 2017, pp. 776--780.

\bibitem{pleiss2017fairness}
G.~Pleiss, M.~Raghavan, F.~Wu, J.~Kleinberg, and K.~Q. Weinberger, ``On
  fairness and calibration,'' \emph{NeurIPS}, vol.~30, 2017.

\bibitem{guo2017calibration}
C.~Guo, G.~Pleiss, Y.~Sun, and K.~Q. Weinberger, ``On calibration of modern
  neural networks,'' in \emph{ICML}.\hskip 1em plus 0.5em minus 0.4em\relax
  PMLR, 2017, pp. 1321--1330.

\bibitem{thulasidasan2019mixup}
S.~Thulasidasan, G.~Chennupati, J.~A. Bilmes, T.~Bhattacharya, and S.~Michalak,
  ``On mixup training: Improved calibration and predictive uncertainty for deep
  neural networks,'' \emph{NeurIPS}, vol.~32, 2019.

\bibitem{minderer2021revisiting}
M.~Minderer, J.~Djolonga, R.~Romijnders, F.~Hubis, X.~Zhai, N.~Houlsby,
  D.~Tran, and M.~Lucic, ``Revisiting the calibration of modern neural
  networks,'' \emph{NeurIPS}, vol.~34, 2021.

\bibitem{fernando2021dynamically}
K.~R.~M. Fernando and C.~P. Tsokos, ``Dynamically weighted balanced loss: class
  imbalanced learning and confidence calibration of deep neural networks,''
  \emph{IEEE Transactions on Neural Networks and Learning Systems}, 2021.

\bibitem{platt1999probabilistic}
J.~Platt \emph{et~al.}, ``Probabilistic outputs for support vector machines and
  comparisons to regularized likelihood methods,'' \emph{Advances in large
  margin classifiers}, vol.~10, no.~3, pp. 61--74, 1999.

\bibitem{zadrozny2001obtaining}
B.~Zadrozny and C.~Elkan, ``Obtaining calibrated probability estimates from
  decision trees and naive bayesian classifiers,'' in \emph{ICML},
  vol.~1.\hskip 1em plus 0.5em minus 0.4em\relax Citeseer, 2001, pp. 609--616.

\bibitem{kull2019beyond}
M.~Kull, M.~Perello~Nieto, M.~K{\"a}ngsepp, T.~Silva~Filho, H.~Song, and
  P.~Flach, ``Beyond temperature scaling: Obtaining well-calibrated multi-class
  probabilities with dirichlet calibration,'' \emph{NeurIPS}, vol.~32, 2019.

\bibitem{rahimi2020intra}
A.~Rahimi, A.~Shaban, C.-A. Cheng, R.~Hartley, and B.~Boots, ``Intra
  order-preserving functions for calibration of multi-class neural networks,''
  \emph{NeurIPS}, vol.~33, pp. 13\,456--13\,467, 2020.

\bibitem{patel2021multi}
K.~Patel, W.~Beluch, B.~Yang, M.~Pfeiffer, and D.~Zhang, ``Multi-class
  uncertainty calibration via mutual information maximization-based binning,''
  in \emph{ICLR}, 2021.

\bibitem{muller2019does}
R.~M{\"u}ller, S.~Kornblith, and G.~E. Hinton, ``When does label smoothing
  help?'' \emph{NeurIPS}, vol.~32, 2019.

\bibitem{mukhoti2020calibrating}
J.~Mukhoti, V.~Kulharia, A.~Sanyal, S.~Golodetz, P.~Torr, and P.~Dokania,
  ``Calibrating deep neural networks using focal loss,'' \emph{NeurIPS},
  vol.~33, pp. 15\,288--15\,299, 2020.

\bibitem{lin2017focal}
T.-Y. Lin, P.~Goyal, R.~Girshick, K.~He, and P.~Doll{\'a}r, ``Focal loss for
  dense object detection,'' in \emph{Proceedings of the IEEE international
  conference on computer vision}, 2017, pp. 2980--2988.

\bibitem{brocker2009reliability}
J.~Br{\"o}cker, ``Reliability, sufficiency, and the decomposition of proper
  scores,'' \emph{Quarterly Journal of the Royal Meteorological Society: A
  journal of the atmospheric sciences, applied meteorology and physical
  oceanography}, vol. 135, no. 643, pp. 1512--1519, 2009.

\bibitem{gal2016dropout}
Y.~Gal and Z.~Ghahramani, ``Dropout as a bayesian approximation: Representing
  model uncertainty in deep learning,'' in \emph{ICML}.\hskip 1em plus 0.5em
  minus 0.4em\relax PMLR, 2016, pp. 1050--1059.

\bibitem{penha2021calibration}
G.~Penha and C.~Hauff, ``On the calibration and uncertainty of neural learning
  to rank models for conversational search,'' in \emph{Proceedings of the 16th
  Conference of the European Chapter of the Association for Computational
  Linguistics: Main Volume}, 2021, pp. 160--170.

\bibitem{cohen2021not}
D.~Cohen, B.~Mitra, O.~Lesota, N.~Rekabsaz, and C.~Eickhoff, ``Not all
  relevance scores are equal: Efficient uncertainty and calibration modeling
  for deep retrieval models,'' in \emph{SIGIR}, 2021, pp. 654--664.

\bibitem{durasov2021masksembles}
N.~Durasov, T.~Bagautdinov, P.~Baque, and P.~Fua, ``Masksembles for uncertainty
  estimation,'' in \emph{Proceedings of the IEEE/CVF Conference on Computer
  Vision and Pattern Recognition}, 2021, pp. 13\,539--13\,548.

\bibitem{lakshminarayanan2017simple}
B.~Lakshminarayanan, A.~Pritzel, and C.~Blundell, ``Simple and scalable
  predictive uncertainty estimation using deep ensembles,'' \emph{NeurIPS},
  vol.~30, 2017.

\bibitem{zhang20mix}
J.~Zhang, B.~Kailkhura, and T.~Y.-J. Han, ``Mix-n-match: Ensemble and
  compositional methods for uncertainty calibration in deep learning,'' in
  \emph{ICML}.\hskip 1em plus 0.5em minus 0.4em\relax PMLR, 2020, pp.
  11\,117--11\,128.

\bibitem{charoenphakdee2021focal}
N.~Charoenphakdee, J.~Vongkulbhisal, N.~Chairatanakul, and M.~Sugiyama, ``On
  focal loss for class-posterior probability estimation: A theoretical
  perspective,'' in \emph{Proceedings of the IEEE/CVF Conference on Computer
  Vision and Pattern Recognition}, 2021, pp. 5202--5211.

\bibitem{liu2020simple}
J.~Liu, Z.~Lin, S.~Padhy, D.~Tran, T.~Bedrax~Weiss, and B.~Lakshminarayanan,
  ``Simple and principled uncertainty estimation with deterministic deep
  learning via distance awareness,'' \emph{NeurIPS}, vol.~33, pp. 7498--7512,
  2020.

\bibitem{tzanetakis2002musical}
G.~Tzanetakis and P.~Cook, ``Musical genre classification of audio signals,''
  \emph{IEEE Transactions on speech and audio processing}, vol.~10, no.~5, pp.
  293--302, 2002.

\bibitem{ajmera2011text}
P.~K. Ajmera, D.~V. Jadhav, and R.~S. Holambe, ``Text-independent speaker
  identification using radon and discrete cosine transforms based features from
  speech spectrogram,'' \emph{Pattern Recognition}, vol.~44, no. 10-11, pp.
  2749--2759, 2011.

\bibitem{liang2018enhancing}
S.~Liang, Y.~Li, and R.~Srikant, ``Enhancing the reliability of
  out-of-distribution image detection in neural networks,'' in
  \emph{International Conference on Learning Representations}, 2018.

\bibitem{szegedy2016rethinking}
C.~Szegedy, V.~Vanhoucke, S.~Ioffe, J.~Shlens, and Z.~Wojna, ``Rethinking the
  inception architecture for computer vision,'' in \emph{Proceedings of the
  IEEE conference on computer vision and pattern recognition}, 2016, pp.
  2818--2826.

\bibitem{huang2017densely}
G.~Huang, Z.~Liu, L.~Van Der~Maaten, and K.~Q. Weinberger, ``Densely connected
  convolutional networks,'' in \emph{Proceedings of the IEEE conference on
  computer vision and pattern recognition}, 2017, pp. 4700--4708.

\bibitem{naeini2015obtaining}
M.~P. Naeini, G.~Cooper, and M.~Hauskrecht, ``Obtaining well calibrated
  probabilities using bayesian binning,'' in \emph{Twenty-Ninth AAAI Conference
  on Artificial Intelligence}, 2015.

\end{thebibliography}


\end{document}